# The impact of design factors of virtual and augmented reality on tertiary students' user experience in Metaverse

**Julius G. Garcia[1], Alvie Simonette Q. Alip[2]**
[1]Faculty of College of Industrial Technology, Technological University of the Philippines, Manila, Philippines
[2]Office of the Vice Chancellor for Academic Affairs, University of the Philippines Open University, Los Banos, Philippines

**ABSTRACT**

The Metaverse is a convergent space integrating virtual reality (VR) and augmented reality (AR) technologies, with market projections rising from $65.5 billion in 2022 to $1.3 trillion by 2030. Despite rapid adoption in education, the specific contributions of visual elements, environmental design, and communication features to user experience (UX) remain underexplored, limiting evidence-based design and resource allocation. This study examined how these design factors in VR and AR environments influence UX in Metaverse platforms. Using a correlational research design, data were collected from 321 purposively sampled tertiary students from engineering and computer science departments across four higher education institutions, all familiar with Metaverse platforms. A structured questionnaire with validated 5-point Likert scales measured UX; visual elements (field of view, resolution, color, complexity, and style); environmental design (interactivity, cohesiveness, naturalness, and complexity); and communication features (text/audio/video tools, avatar interaction, and spatial audio). Multiple regression analysis assessed the independent and combined effects of these factors on UX. Results show that environmental design ($\beta$=0.424, $p$<0.001) and visual elements ($\beta$=0.298, $p$<0.001) are the primary predictors of UX, jointly explaining 47.3% of its variance, with a strong correlation between them ($r$=0.792). Communication features did not significantly influence UX ($\beta$=0.053, $p$=0.196), challenging assumptions about virtual connectedness and indicating that communication design should better support social interaction. These findings highlight the importance of well-designed, interactive virtual spaces and suggest that developers and educational institutions prioritize environmental design to advance immersive learning and broader Metaverse applications.

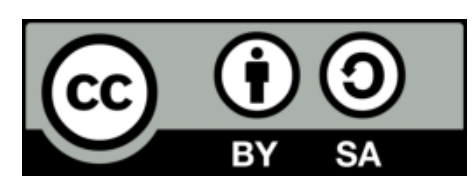

***Corresponding Author:***

Julius G. Garcia
Faculty of College of Industrial Technology, Technological University of the Philippines
Ayala Blvd., Ermita, Manila, Philippines
Email: julius.tim.garcia@gmail.com

## 1. INTRODUCTION

Digitally augmented and physically persistent virtual reality comprise the Metaverse, a shared environment. It encouraged immersive technology. Bloomberg Intelligence expects the projected growth to reach about $140 billion by 2025, driven largely by token-based transactions, which are expected to grow at over 60% per year [1]. Virtual worlds are developing quickly, making user experience (UX) more vital. The new technology has driven virtual reality (VR)/augmented reality (AR) adoption. Hardware improvements affect weight, display resolution, and motion tracking. Parallel software enables realistic graphics, improved

user interfaces, and more interaction paradigms. Companies sought virtual cooperation, education, and pleasure during the pandemic.

Meta (previously Facebook), Microsoft, Apple, and Google have invested heavily in Metaverse technology, demonstrating their optimism for immersive digital worlds. Metaverse VR and AR remain difficult despite advances. Problems with UX VR/AR sessions cause cybersickness, ocular fatigue, and exhaustion [2]. Limited field of view (FOV), resolution, and latency hamper virtual presence [3], [4]. Virtual world users detest poor graphics, engagement, and navigation [5], [6]. Problems include virtual communication. Collaboration and social interaction in Metaverse apps hinder natural communication. Texting is less classy than face chat. Several VR worlds have communication issues. Avatar-based nonverbal communication may miscommunicate or reduce social presence with simpler body language and facial expressions [7].

The environment greatly affects VR/AR. VR must balance vision, layout, navigation, and complexity. Complex environments might mislead consumers, but simple ones can bore them. Style, resolution, color palettes, contrast ratios, and vision affect VR interaction. Poor execution can cause eye strain, immersion loss, and a lack of appeal. Demography and application affect graphic fidelity and art [8]. The relevance and interdependencies of UX components explain the complexity of Metaverse VR/AR applications. Few quantitative studies have examined how visuals, environment, and communication affect UX. Immersive digital experience developers must close this knowledge gap to prioritize design and allocate resources. The effects of visual aspects, environmental design, and communication features on UX are unclear. Most research examines these factors individually, leaving a gap in our understanding and making it hard for designers and institutions to allocate resources or set design objectives, especially for VR/AR Metaverse applications.

This study empirically evaluates key educational Metaverse design elements to fill that need. A hierarchical design approach is proposed to examine how each element affects tertiary students' immersive learning. Researchers identify the most significant factors to help designers allocate time, effort, and cash. It offers tips for developing Metaverse applications in education, entertainment, healthcare, and professional collaboration. This study addresses the following research questions (RQ):
- Do visual elements, environmental design, and communication features independently predict UX in VR/AR Metaverse? (RQ1)
- What is the hierarchical order of design factors based on their relative contribution to UX variance among tertiary students? (RQ2)
- How do visual elements, environmental design, and communication features interact synergistically to shape UX in Metaverse contexts? (RQ3)

## 2. LITERATURE REVIEW

### 2.1. Visual elements in VR and AR

Graphics make VR/AR fun. FOV affects immersion and look. Wider FOVs make you feel more immersed and less like a window [8], [9]. Although a wide FOV can create visual distortion and discomfort, it may require adjustments in application and hardware. Display resolution and visual fidelity affect immersion. High resolution decreases pixilation artifacts, increasing photo quality and realism [5], [6]. Representing materials, lighting, and shadows accurately. Space-based reasoning and visual assessment are more fun in high-fidelity VR. Color theory and contrast management make things visible and comfortable. Colors affect moods and the surroundings. Virtual residents can move and interact easier with contrast [10], [11]. Insufficient color and contrast might fatigue, confuse, and dissatisfy consumers. Colors like reds, oranges, and yellows thrill and awaken, while blues, greens, and purples soothe. This allows designers to intentionally affect people. For the best experience, improve visual complexity. Visual intricacy might make information digest and decision-making difficult.

A lack of environmental detail bores and disengages. The task and user proficiency determine visual complexity. To comprehend the context, experienced users prefer greater visual detail, whereas new users prefer simpler surroundings that do not require much thought [12]. Users can feel and think differently depending on how something looks and seems real. Simulations are more lifelike with photorealistic visuals.

According to Eswaran and Bahubalendruni [13], abstract or stylized visuals increase artistic flexibility while decreasing hardware performance, thereby preserving user attention. Recent research demonstrates that the application environment and target demographics greatly affect the suitability of visual style. To help students apply what they learn in real life, professional training simulations need realistic graphics, but kid-friendly instructional apps use cartoon imagery.

### 2.2. Environmental design principles

Environmental design organizes, interacts, and integrates virtual space. For interest, virtual environments must be interactive. Altering products and engaging with the environment encourages user participation and autonomy. Research shows that interactive settings make people happier, help them remember tasks, and teach better than watching. Interactive learning encourages experimentation and engagement with virtual content. Virtual worlds that feel real and cohesive make people feel more immersed. Cohesive environments complement rather than compete [14].

Architecture, lighting, and physical properties like gravity and material response may make users doubt what they see or feel. Interaction model, audio design, and tactile feedback create a cohesive environment. Design must consider how natural VR is and how AR integrates with everything. Natural VR's familiar spatial relationships, intuitive bodily reactions, and environmental patterns help users adapt and reduce cognitive load [7], [15].

Natural situations are easier to navigate than abstract ones. VR aspects in real-world environments do not interfere with AR vision or actions. AR requires perfect spatial registration, realistic lighting, and occlusion handling to match actual and virtual objects. Like visual complexity, environmental complexity is optimized. Complex environments with many interactive features, intricate spatial layouts, and advanced navigation paths may make it difficult for users to complete tasks. Simple settings, devoid of detail or engagement, may dull users and cause them to stop using the program [12]. Application goals, user skills, and job requisites determine an environment's complexity. Training sims become more sophisticated as users improve by adding environmental features. Entertainment apps offer basic difficulties to keep you engaged and learning.

### 2.3. Communication features in virtual environments

Communication enables Metaverse socialization. Metaverse socialization requires communication. Although restricted, text, audio, and video chat are crucial in virtual settings. Texts are straightforward and consistent, but they lack tone, tempo, and loudness, which indicate emotion and intent. Vocal communication preserves voice elements but can degrade audio quality, distort spatial audio, or blend with virtual worlds. Virtual embodiment makes avatar-based communication like a voiceless conversation.

In avatar-mediated communication research, avatar quality, user identification, and social presence are difficult. Avatars with realistic facial expressions and body language enhance social presence and nuanced communication [7], [15]. If avatar realism is uncomfortable, uncanny valley effects, discomfort from near realistic but inadequate human representations, can hamper communication. Stylized avatars avoid the uncanny valley but may lack complex social expression.

Spatial audio improves virtual communication. Multi-user talks benefit from spatial audio systems' sound directionality and distance attenuation. Spatial audio boosts speaker recognition, conversational overlap, and social presence. Advanced spatial audio systems are more realistic with environmental acoustics, material-based sound reflection, and occlusion. Haptic feedback, brain-computer connections, and AI-backed communication enable virtual involvement. Haptic technologies provide virtual handshakes, object passing, and intimacy. Early research suggests haptic communication promotes social connection and task performance in virtual collaboration. Hardware, cost, and user acceptance limit haptic technology.

### 2.4. Integrated UX models

Modern UX research increasingly acknowledges that visual components, environmental design, and communication features function interdependently rather than in isolation. Integrated UX models suggest that these factors interact in a way that makes their combined effects stronger than each would have on its own. For example, high-quality visual elements unify the setting by providing a consistent aesthetic foundation [8], [11]. Good environmental design enhances communication by creating the right social settings and spatial arrangements for interaction [7]. Even while people are starting to understand how different UX factors work together, there is not much research that measures how much each of these categories, visual, environmental, and communication, adds to the overall UX.

Most current research focuses on individual components in isolation or investigates pairwise correlations without thorough multivariate analysis. VR/AR development teams struggle to make evidence-based design decisions and maximize resource allocation due to a lack of research [13]. Knowing which UX factors matter most helps you design development and quality improvement programs [13].

### 2.5. Conceptual framework

UX is the dependent variable, and visual elements, environmental design, and communication are independent factors. Features vary independently. Immersive technology works best when its components are interwoven, according to interaction design theory and presence research. Field of vision, resolution, fidelity, color, contrast, complexity, style, and realistic calibration are visual qualities. These are essential for VR visual processing and perception. Environment design factors include interaction, cohesiveness, naturalness (VR/AR), and complexity. They affect spatial awareness, mobility, and interaction. Text, audio, video, avatar-based interaction, current technology, and spatial fidelity are communication variables. These facilitate internet interaction. The model shows that each UX component affects interaction differently. Identifying these relationships helps developers plan and enhance virtual worlds. Figure 1 shows how three independent variables affect UX.

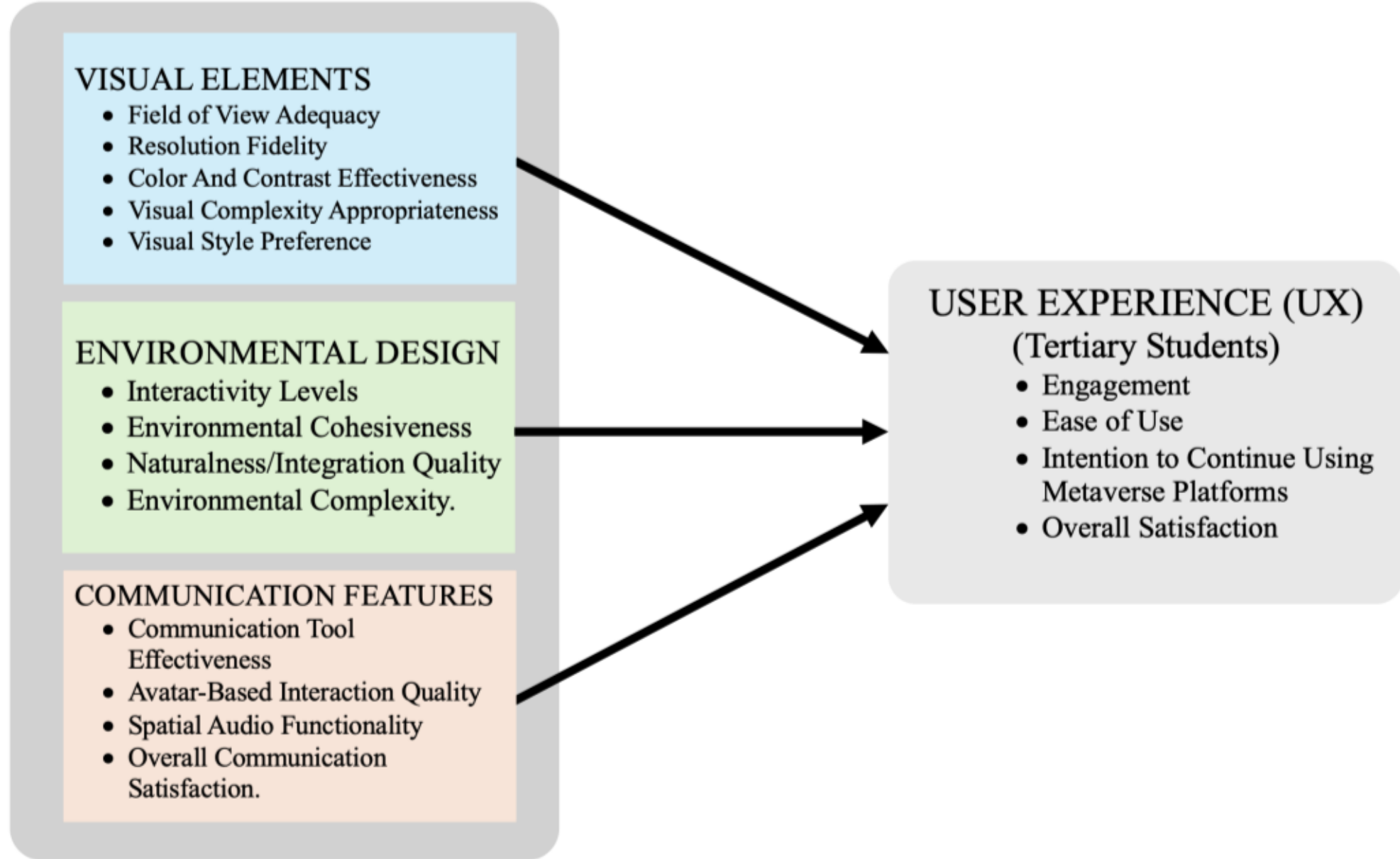

Figure 1. Conceptual framework of the study

## 3. METHOD

### 3.1. Research design

A quantitative correlational design using multiple regression was employed to examine how different variables affect users' Metaverse VR/AR experience. This multivariate approach evaluates all variables at once, provides standardized coefficients (β) that show each factor's independent influence, and uses regression diagnostics ($R^2$, adjusted $R^2$, residual analysis) to assess model performance. Structural equation modeling and analysis of variance (ANOVA) were not used because current theory does not support strong causal assumptions, and converting continuous design indicators into categories would reduce data richness and statistical power.

### 3.2. Participants and sampling

Most of the 321 surveyed students from four higher education institutions were engineering and computer science majors. Purposive sampling was used to recruit participants who were Metaverse-savvy. They were required to be tech-savvy and interested in learning about the Metaverse. All participants received hands-on training in navigating and communicating on the Metaverse platform before data collection. This ensured a uniform technical understanding and reduced the risk that unfamiliarity with the technology would affect their evaluations. Multiple regression was conducted. This sample size exceeded the recommended 15 to 20 participants per predictor, providing statistical power above 0.95.

### 3.3. Instrumentation: development and validation process

The four primary constructs, as presented in Table 1, were assessed using a structured questionnaire with 5-point Likert scales (1=strongly disagree, 5=strongly agree), adapted and combined from established VR and AR UX scales based on immersive experience frameworks, presence theory, and the technology acceptance model (TAM) to capture the technical and experiential impact of each construct on Metaverse VR and AR use. The instrument was validated by 10 VR and AR technology and human-computer interaction experts and pilot-tested with 50 participants to establish internal consistency reliability for all subscales (Cronbach's alpha $\alpha>0.75$), and content validity was established through identifying 68 items from validated instruments and interviews, expert rating of each item on a 4-point scale leading to revision or removal of items with a content validity index (CVI) of 0.78 and a final set of 51 items, and refinement of wording using interview data. A four-factor structure was extracted using principal axis factoring with oblique rotation, explaining 68.4% of the variance, with high reliability, internal consistency, and inter-item correlations (Cronbach's alpha ranging from α=0.85 to α=0.93), and two-week test–retest data from 25 subjects yielded ICCs of 0.82 to 0.89. Strong relationships with the system usability scale (SUS, r=0.73) and presence questionnaire (r=0.68) supported convergent validity, while moderate correlations among the four factors supported discriminant validity and confirmed the distinctiveness of each factor, indicating that the instrument is dependable and robust for evaluating UX in educational Metaverse use.

Table 1. Study variables

| Construct (variable type) | Source | Subdimensions |
|---|---|---|
| UX (dependent variable) | Laugwitz *et al*. [16] | 1. Overall satisfaction |
| | Jennett *et al*. [17] | 2. Engagement |
| | | 3. Ease of use |
| | | 4. Continuance intention |
| Visual elements | McMahan *et al*. [6] | 1. FOV adequacy |
| (independent variable) | Ragan *et al*. [8] | 2. Resolution and visual fidelity |
| | Gonçalves *et al*. [18] | 3. Color and contrast effectiveness |
| | | 4. Visual complexity appropriateness |
| | | 5. Visual style and realism |
| Environmental design | Slater and Sanchez-Vives [4] | 1. Interactivity levels |
| (independent variable) | Bailenson *et al*. [7] | 2. Environmental cohesiveness |
| | Liu *et al*. [12] | 3. Naturalness and integration quality |
| | | 4. Environmental complexity |
| Communication features | Moreira *et al*. [19] | 1. Communication tool effectiveness |
| (independent variable) | Sandi and Kalfin [20] | 2. Avatar-based interaction quality |
| | Yousefdeh and Oyelere [21] | 3. Spatial audio functionality |
| | Li *et al*. [22] | 4. Overall communication satisfaction |

### 3.4. Data collection procedures

The data-gathering process unfolded in three main stages. First, participants were briefed on the study and asked to provide their informed consent. They then participated in a set of standardized activities on a Metaverse platform, providing them with sufficient direct experience to make an informed decision in the next phase. After their time on the platform, they completed a structured questionnaire designed to capture their current impressions and experiences. Data collection spanned a six-week period to accommodate participants' varying schedules and to achieve a sufficient sample size.

### 3.5. Data analysis

Data analysis used descriptive and inferential statistics. Means, standard deviations, and frequency distributions described sample demographics and variable distributions. Pearson correlation coefficients were used to examine all bivariate correlations. A multiple linear regression analysis examined how visual, contextual, and communicative factors impact user enjoyment. We validated the approach using residual analysis, multicollinearity evaluation (variance inflation factor (VIF) values), and assumption verification. We used α=0.05 to evaluate statistical significance. All statistical studies used SPSS 28.0.

## 4. RESULTS

### 4.1. Descriptive statistics

Table 2 displays descriptive statistics for all study variables. All factors used 5-point Likert scales (1=strongly disagree, 5=strongly agree). Visual elements had the highest mean rating (M=3.92, SD=0.55), indicating that most participants considered all Metaverse platforms had good visual quality. UX achieved the mean value of 3.81 (SD=0.67), which indicated that the majority held positive experiences. Environment design held the lowest mean value among the three predictor variables, 3.77 (SD=0.55) but still high. Communication features varied the widest (M=3.82, SD=0.74), which indicated a broader range in participant opinions regarding communication capacity.

### 4.2. Correlation analysis

The correlation matrix for all research variables shown in Table 3. Visual elements (r=0.634, p<0.001) and environment design (r=0.661, p<0.001) strongly connect with UX. This shows this design variables considerably impacted user happiness. Environment design was more UX-related than visual elements, possibly showing regression. Communication functionality had a weak, non-significant relationship with UX (r=0.069, p=0.107). This surprising result puts into question social virtual communication's importance and deserves further study. Visual elements and environment design showed substantial connection (r=0.792, p<0.001), indicating shared variance.

The strong link between visual and environmental factors suggests a connection or benefit. Due to substantial correlation, multicollinearity may complicate regression analysis. Additional VIF tests confirmed no significant influence of multicollinearity on regression results (all VIF<3.0). Communication aspects were unrelated to visual elements (r=0.010, p=0.854) and environment design (r=0.032, p=0.572), showing they work independently. Communication qualities seem to be separate from visual-environmental design.

Table 2. Descriptive statistics of study variables

| Variable | Mean | Std. Deviation | N |
|---|---|---|---|
| UX | 3.8107 | 0.6684 | 321 |
| Visual elements | 3.9195 | 0.5528 | 321 |
| Environment design | 3.7686 | 0.5488 | 321 |
| Communication features | 3.8185 | 0.7353 | 321 |

Table 3. Correlation matrix among study variables

| Variable | UX | Visual | Environment | Communication |
|---|---|---|---|---|
| UX | 1.000 | 0.634*** | 0.661*** | 0.069 |
| Visual | 0.634*** | 1.000 | 0.792*** | 0.010 |
| Environment | 0.661*** | 0.792*** | 1.000 | 0.032 |
| Communication | 0.069 | 0.010 | 0.032 | 1.000 |

Note. ***p<0.001

### 4.3. Regression model fitness

Table 4 illustrates the regression model's overall fit. Multiple correlation coefficient (R=0.688) reveals that predictor factors affect UX overall. The coefficient of determination ($R^2$=0.473) indicates that the three predictor factors explain 47.3% of the UX difference. For behavioral research, where numerous unmeasured factors effect outcomes, this is a high percentage. The modified $R^2$ (0.468) considers the sample size and number of predictors. The $R^2$ has somewhat decreased from the unadjusted value. This little disparity shows that the model avoids overfitting and that the predictors provide significant explanatory power rather than sample-specific variance. For regression, the residuals must have no substantial autocorrelation, and the Durbin-Watson statistic (2.071) is close to 2.0.

### 4.4. ANOVA results

ANOVA results in Table 5 illustrate the regression model's overall importance. The significant F-statistic (F(3, 317)=94.783, p<0.001) confirms that the regression model fully accounts for UX variance. This means that visual elements, environment design, and communication features together have predictive value that exceeds random predictions. The significant ANOVA result supports individual predictor coefficient analysis to find the factors affecting observed relationships.

Table 4. Regression model fitness summary

| Model | R | R square | Adjusted R square | Std. Error |
|---|---|---|---|---|
| 1 | 0.688 | 0.473 | 0.468 | 0.48754 |

Table 5. ANOVA: overall significance of regression model

| Source | Sum of squares | df | Mean square | F |
|---|---|---|---|---|
| Regression | 67.590 | 3 | 22.530 | 94.783*** |
| Residual | 75.351 | 317 | 0.238 | |
| Total | 142.940 | 320 | | |

Note. ***p<0.001

### 4.5. Regression coefficients

Table 6 presents regression coefficients showing how much each predictor affects UX alone. UX was most influenced by environment design (β=0.424, t=6.349, p<0.001). Interactivity, cohesiveness, naturalness, and suitable complexity had the greatest impact on user satisfaction. This suggests that development should prioritize well-designed, interconnected, and engaging virtual environments.

Table 6. Regression coefficients analysis

| Predictor | B | SE | β | t | p | VIF |
|---|---|---|---|---|---|---|
| (Constant) | 0.270 | 0.245 | | 1.101 | 0.272 | |
| Visual elements | 0.360 | 0.081 | 0.298 | 4.462 | <0.001*** | 2.689 |
| Environment design | 0.516 | 0.081 | 0.424 | 6.349 | <0.001*** | 2.677 |
| Communication features | 0.048 | 0.037 | 0.053 | 1.295 | 0.196 | 1.001 |

Note. ***p<0.001

Visual aspects have a secondary impact on UX (β=0.298, t=4.462, p<0.001). Environmental design matters more than aesthetics. This suggests that high-quality graphics function best in well-designed environments rather than as indicators of quality. Communication Features did not significantly predict UX (β=0.053, t=1.295, p=0.196). This null result contradicts social virtual communication wisdom. The non-significant result suggests that basic communication capability may be sufficient for general user satisfaction, or that communication features influence experiential characteristics such as social presence and collaborative effectiveness, which overall UX measurements fail to capture. The regression equation as in (1):

$$UX = 0.270 + 0.360(Visual) + 0.516(Environment) + 0.048(Communication) \quad (1)$$

This equation estimates UX scores using visual, environmental, and communication factors. Since the communication coefficient is not significant, this section does not provide much practical prediction value.

### 4.6. Residual diagnostics

Residual analysis confirmed regression. The mean residual of 0.000 suggests impartial forecasts. Standardized residuals (-2.713 to 2.302) were within the acceptable ±3 level, indicating no significant outliers affecting results. The standard deviation of residuals (0.485) is close to the standard error of estimate (0.488), indicating prediction accuracy across the anticipated values. Histogram and P-P plots showed almost normal residual distributions, supporting the hypothesis testing validity of the normality assumption. Scatterplot patterns showed random dispersion without systematic patterns, indicating homoscedasticity (constant variance).

## 5. DISCUSSION

### 5.1. The primacy of environment design

Environment design is paramount in the creation of virtual worlds as it largely affects the experience (β=0.424). This only validates theories that emphasize presence and spatial immersion for profound experiences [15]. This also reveals that users are fonder of useful, coherent, and interesting settings compared to other design aspects. Environmental attributes can explain the dominance. Environmental interactivity enables people to explore and modify virtual environments. Interactivity in virtual worlds increases the curiosity and dynamic interaction of users and affects behavioral and cognitive involvement. User interaction elevates interactive learning and experiences. Immersive VR environments facilitating user control augment immersion, intrinsic motivation, and well-being, and confirm that people need agency. Environmental consistency in the domains of virtual worlds improves immersion and cognitive dissonance. Smooth ambient enables people to experience virtual entertainment. Recent research indicates that coherent settings with seamless spatial, visual, and interactive aspects yield homogeneous virtual environments that heighten people's presence and reduce distractions [10].

Inconsistent environmental attributes reduce immersion and highlight the fabricated nature of learning Metaverse locations, reducing the enjoyment in users. High environment design coefficients calibrate users with coherence. Environmental naturalness and spatial arrangement should be given more attention and focus. Proximity to points of focus and considerate spatial arrangement aid learning, attention, and the VR classroom experience. User interaction and treatment efficacy are enhanced in high-fidelity virtual worlds with naturalistic acoustic landscapes and appropriate environmental acoustics. These guidelines assert that environmental design must be taken into account by VR/AR developers. It is recommended to have a strong interaction systems with natural, intuitive mechanisms, visual, auditory, and interactive consistency across virtual worlds, environmental complexity with thoughtful balance to counteract engagement and cognitive load, and intensive usability testing for the detection of environmental incoherency or navigation Development teams should design the environment early and in iterative refinement in order to produce coherent spatial experiences in which the user is encouraged and cued to take agency and exploration.

### 5.2. Higher education implications for Metaverse learning design

These results substantially influence how higher education institutions should use the Metaverse. When investing in Metaverse-based learning environments, institutions should prioritize supporting teaching and learning over aesthetics. Place interactive features at the center of virtual classes and labs. Experiments, object manipulation, and environmental learning should be possible for students. Rich simulation settings where students may alter variables, monitor changes, and build comprehension through direct engagement should replace virtual lecture halls that simply mimic real-world rooms for one-way content delivery.

These findings suggest that educational institutions building educational Metaverse platforms should carefully consider communication spending. Schools should prioritize simple, effective communication tools over complex systems with many functions, unless specific educational activities require them. Simple text, voice, and avatar interactions may be enough for general learning. This would free up resources for larger environmental and aesthetic improvements.

Some schools must invest more in communication in a certain way. Cooperative learning activities, including group projects, peer review, and discussion seminars, may utilize greater communication. Spatial audio, expressive avatars, and collaborative tools like shared whiteboards and co-editing allow people to communicate and change groupings. Instead of adding features to everything, invest in what makes sense. Educational designers should conduct activity-specific analysis to determine communication needs. Lectures, demonstrations, and one-on-one practice do not require good communication skills. Breakout talks, problem-solving, and teaching each other require more clarity. This requires tiered communication systems that only enable sophisticated functionality in certain social circumstances. This will improve performance and reduce costs. Higher education institutions should also consider how student body size affects communication. People who grew up with computers, use social media, and play online games may be able to survive without communication tools by being innovative and obeying the rules. Students unfamiliar with mediated communication may need more intuitive and feature-rich solutions to attain similar interaction quality. When communicating with a certain student group, consider this demography. The study implies that communication spending should be based on specific instructional goals rather than what we think is essential. Before investing heavily in communication technology, higher education institutions should determine if better tools improve learning, student satisfaction, and collaboration.

### 5.3. The significant but secondary role of visual elements

Visual aspects enhanced the UX but were less influential than other features ($\beta$=0.298). Visual quality and design hierarchy are stressed for high fidelity in VR systems. FOV, resolution, color utilization, and realism enhance the experience but cannot compensate for flawed design. Display quality and visual fidelity significantly affect presence, immersion, and learning. High visual element coefficient agrees. Resolution, pixel area, and visual quality alleviate screen-door effect, visual fatigue, and immersion in advanced VR/AR displays. FOV enhances spatial awareness and immersion due to peripheral vision [5], [8]. Color precision and sharpness in dynamic illumination can enhance visual perception up to 41% and enhance user comfort and satisfaction.

Recent research stresses visual reality calibration. Photorealistic self-avatars were preferred in VR users, while first impressions and the choice of context favored stylization over photorealism. High-fidelity surroundings enhance immersion and interaction with functional fidelity. Environmental realism can alter context-dependent forgetting and source-monitoring and hinder visual fidelity and presence. The high correlation (r=0.792) from visual elements and environment design suggests they interact together. High visual fidelity enhances environmental realism, while low visual graphics harm spatial arrangements. Integration of visual and environmental design ensures spatial aesthetics consistency. Visual quality should be good, but developers should remember that perfection does not improve UX. Arranging resources should show that visually matter but are not the main focus. Adjust display parameters (FOV, resolution, and refresh rate) to optimize performance and visual fidelity within hardware constraints. Color theory improves mood and function. Visually verify all virtual areas for aesthetic uniformity.

### 5.4. Challenging social VR assumptions and context-dependent effects of communication attributes

This study found no significant impact of communication elements on UX ($\beta$=0.053, p=0.196). It contradicts the key premise of social VR research that Metaverse systems seek social connectivity [7]. This null conclusion necessitates a theoretical reevaluation and affects institutions and developers' educational investment decisions in the Metaverse. This contradicts Meta and Microsoft Mesh, which emphasize social presence, avatars, and collaborative capabilities over traditional online learning. Communication quality may be more of a hygiene aspect in higher education: when tools break, students are unhappy, but once communication is satisfactory, further enhancements have little impact on UX.

This pattern can be explained by several possibilities. The threshold satisfaction hypothesis states that present platforms provide communication tools sufficient for most students, and adding complexity (e.g., high-fidelity facial animation, precise gesture tracking, haptics) yields diminishing returns. Hygiene factors are not true motivators, according to two-factor motivation theory [23].

The context-specificity hypothesis states that communication importance depends on use case. Communication quality likely affects social presence and coherence in virtual events, collaborative studios, and peer discussions. It matters less in content-focused environments (lectures, solitary simulations, and individual practice). This study combined instructional applications that may have had favorable effects in social contexts but not in lonely ones.

The measurement specificity hypothesis suggests that our measures do not align with the communication constructs. UX scales generally measure usability, engagement, and satisfaction. Standard UX tools only partially cover social presence, copresence, connection, and collaboration effectiveness, but communication quality directly affects them. Communication can greatly affect social experience without changing overall UX scores, according to recent research [24].

The sufficient functionality hypothesis states that simple instruments can efficiently support ordinary educational communication, such as asking questions, getting feedback, organizing group work, and facilitating class discussions. Students often work around platform limitations to interact well. Advanced features such as realistic avatars, complex nonverbal systems, and high-quality spatial audio may not be useful.

These suggest a more complex view of educational Metaverse communication. Advanced communication features can improve the higher education UX in some contexts and purposes, but not always. Institutions may benefit more from robust, reliable fundamental communication and from allocating resources to other Metaverse aspects that motivate rather than just serve as hygiene for design and investment decisions.

### 5.5. Synergistic relationships among design elements

Visual elements and environmental design are strongly correlated (r=0.792), indicating that design components are interdependent. These features improve UX together due to this strong correlation. Recently conducted integrated design research indicates many synergies. Visuals match the ambient style. Lighting, colors, and styles alter the surroundings. New fidelity evaluation research emphasizes synchronization of visual systems and contextual design to improve experience. Venues with adequate lighting, a clear layout, and distinct focus areas enhance the viewing experience and draw attention. Vision and environment affect how we think and process information. Simple pictures may under stimulate thinking, whereas complex ones may overstimulate. Recent immersion learning study reveals that the optimum experience balances visual and contextual complexity, depending on task and skill level [25].

Cues aid thinking in social situations [22]. Immersive virtual realities that use visual and environmental features to enhance embodied involvement boost intrinsic motivation and learning. Avatars, spatial arrangement, movement, and visual quality affect concentration, usability, presence, and learning. Integration influences experience quality. AR/VR museum exhibits reveal that immersive technology works best with authentic graphics, a sense of place, and interactive design [13], [26]. Integrated and comprehensive designs work best, according to the study. Development teams should not treat visual and environmental design as separate issues that require different solutions. Visual, environment, and interaction designers must collaborate during development. Visual and environmental design evaluations and iterative testing must demonstrate consistency and cooperation. Metaverse UX design best practices emphasize easy-to-use visual and spatial experiences that cover a lot of territory [27], [28].

### 5.6. Theoretical contributions and advancement of virtual experience architecture

This paper advances virtual experience architectural theory in various ways. It was the first to scientifically quantify the hierarchy of design factors using a simultaneous multivariate method rather than paired comparisons or individual factors. The study establishes a priority structure based on environmental design ($\beta$=0.424), visual elements ($\beta$=0.298), and communication features ($\beta$=0.053, non-significant), providing evidence-based information not available in previous theoretical models.

This hierarchy adds subtlety to presence theory by showing that spatial presence, affected by environmental elements, affects user pleasure more than perceptual or social presence, which are visual or communicative characteristics. Slater and Sanchez-Vives [4] believed environmental elements determine presence, but their model did not evaluate strengths. We provide empirical evidence and explain how these components interact in practice to fill that gap.

The study clarifies which design qualities matter most for perceived ease of use and perceived utility, the two main drivers of technology acceptance in these frameworks and extends the TAM for VR. Environmental design promotes spatial usability—including intuitive navigation, consistent interactions, and adequate complexity—which underpins perceived ease of use, with visual quality reinforcing it. The deeper understanding enables targeted TAM deployment in immersive technology scenarios.

The substantial unexplained variance (52.7%) indicates model flaws, providing theoretical insight. Traditional VR and AR frameworks concentrate visual, environmental, and communicative features above human diversity, social dynamics, and context. Individual technology self-efficacy, personality, and past experience; social community quality, relationship development, and collaborative processes; and contextual application purpose, task demands, and usage patterns are understudied. The fact that conventional design factors account for less than half of UX variance means theories must include them.

Communication features contradict several social VR theory principles. Metaverse platforms are often touted for its social connectivity, but our research shows that it does not necessarily improve user satisfaction. Its impact depends on context, users, and results. This shows that more specific theory is needed to identify social and task-focused programs, socially driven and task-oriented users, and overall enjoyment and narrower social results. Future theoretical models should explicitly indicate these boundary limits instead of treating communication equally relevant.

Finally, the strong visual-environmental relationship (r=0.792) strengthens integrated design theory. The substances operate synergistically, not additively. This proposes comprehensive design that aligns design domains rather than maximizing individual item. It supports calls for interdisciplinary development teams and integrated design techniques to reduce visual-environmental mismatches and maximize mutual reinforcement and augmentation.

### 5.7. Practical implications for higher education Metaverse development

These findings provide actionable guidance for higher education institutions and educational technology developers implementing Metaverse learning environments across strategic, operational, and tactical decision domains.

#### 5.7.1. Strategic resource allocation

Funding should be allotted more to environmental design (42.4%) than to visual refinement (29.8%) or context-dependent communication (5.3%). Unless some lessons require students to work together, environmental design should get 60–70% of the resources for design and development, visuals should get 25–30%, and communication should get 5–10%. When designing for the environment, focus on interactive simulation technologies that let people touch and try things out. Space that is logically ordered makes it easier to get around. Logical material flow from how things interact in the real world makes learning easier and less tiring for the brain. Adaptive complexity solutions for different levels of learners take into account important factors that affect user satisfaction.

When investing in visual design, focus on making text easy to read and objects easy to recognize, using consistent aesthetic frameworks that keep the environment cohesive, visual complexity that matches the learner's abilities and the task's demands, and lighting and color systems that help with wayfinding and directing attention over aesthetics. Visual design puts environmental goals ahead of quality. Situations should determine how much money to spend on communication features. There should be little investment in lectures, simulations, and independent practice; moderate investment in study groups, question and answer sessions, and peer feedback; and high investment in group projects, collaborative design, and peer teaching. Instead of assuming that all features are advanced, communication ability should be based on activity-specific analysis.

#### 5.7.2. Operational development processes

Development should start with environmental structures before visual or communicative features. First build spatial layouts, navigation algorithms, interaction mechanics, and environmental logic, then visual refinement and communication tools. This cycle guarantees visual elements support environmental organization and avoids costly redesign when visual assets conflict with spatial needs. In quality assurance, environmental usability testing should precede visual quality assessment. Testing should evaluate navigation and wayfinding, spatial awareness and orientation, interactions, action–consequence relationships, environment consistency and logic, and complexity appropriate for target learners. Visual and communication testing should evaluate how well these elements serve the environmental goal.

Spatial designers, level designers, environmental psychologists, and instructional designers with spatial experience should be team members, not consultants. Software developers, graphic designers, and environmentalists must collaborate for integrated development. As needed, communication specialists should assist with social apps. Modern industry practice and research reinforce these conclusions and help VR/AR development teams make various decisions [27]. For coherence, interaction dynamics, and spatial configuration, environmental design requires time and money. Environment should be valued over visual quality, which should be underfunded. Communication quality assurance may be less important without social interaction apps.

Spatial planning and interaction models before visual execution increase development sequence order. Research on Metaverse development includes contextual scaffolding for visual and interface iteration [28]. Environment should influence visual design to match space. Early social feature testing should minimize communication until contextual and visual cues are established. Quality assurance and testing should prioritize environmental usability and coherence. Tests include navigation, interface usability, spatial awareness, and environmental consistency. Current VR immersion experience models evaluate pragmatic (usability, functionality), hedonic (beauty, novelty), and contextual appropriateness [24]. Consider whether visuals aid or hinder environmental aims when measuring quality.

Companies should include environmental designers in development teams to satisfy team and skill demands. Software engineers and spatial designers (architecture, level design, environmental psychology) are essential. Not most people need visual design skills. Telecommunications experts may be ignored until social media is established. These guidelines are Metaverse UX design best practices. Industry standards need simple navigation, clear information, and well-balanced motion and animation to avoid discomfort [27]. Strategizing artificial intelligence (AI)-driven adaptive systems, individualized recommendations, and enhanced engagement mechanisms increases environmental and visual quality [14], [29].

#### 5.7.3. Tactical design decisions

Design choices should reflect hierarchy. Virtual classrooms should provide lecture, discussion, and breakout sections, interactive components that let students move things about and try things out, and simple access to activity areas. No need for photorealism, but visual quality should be legible and consistent. Communication should involve dependable text and voice; however, avatar systems may not be necessary.

Make virtual labs with realistic interaction mechanics for genuine experiments, environmental causality for inquiry-based learning, and the correct level of challenge for each learner. Though not photorealistic, visual representation should accurately depict relationships and moods. Communication should make teamwork easier, but it should not improve socializing.

When working on a project together, employ shared workspaces that make it easy to move objects, organize shared resources clearly, and facilitate parallel and coordinated work. Visual consistency should improve collaboration, and spatial audio and expressive avatars should be funded. Provide pleasant meeting spaces, natural conversational distance and posture, and easy-to-interact environmental characteristics for social learning. As teaching emphasizes social presence, visual and communication technologies demand additional investment.

#### 5.7.4. Assessment and continuous improvement

Higher education institutions should evaluate their environmental usability support, visual learning support, and Metaverse communication sufficiency. Students' feedback forms can score environmental satisfaction, visual quality support, and communication support functionality using the proposed framework's scales. Comparisons of Metaverse deployments will assist institutions determine the relative importance of factors in specific circumstances and create design recommendations for specific uses and audiences. Over time, evidence will refine concepts and identify boundary situations that require tailored deployments.

### 5.8. Contextualization within existing literature

These findings strengthen immersive technology literature. Presence theory emphasizes spatial involvement and interactivity for virtual encounters [4]. Recent meta-analyses show that immersive VR settings with strong environmental underpinnings offer slight but significant advantages over non-immersive ones [5], [30], [31]. The strong visual element effect supports previous display quality studies [5], [8], and new displays increase immersion and reduce simulator sickness. The communication outcome defies research on social presence and group-oriented behavior in virtual environments [7], and theoretical models need to be revised, or communication outcomes are more multidimensional than previously postulated. Recent research accounts for this discrepancy. COVID-19 social VR applications research demonstrated complex relations between communication qualities, social presence, and mental outcomes defined by the application environment and the users' needs. Virtual agents and avatars in immersive settings affect communication efficiency beyond technological sophistication, due to design, behavioral veracity, and environmental congruence. The research provides quantitative coefficients allowing desirable comparisons, improving understanding. Previous research has focused on single constituents, and only a limited number of studies have examined multiple design variables. This approach, based on multiple regression with standardized coefficients, enables evidence-based prioritization and differs from previous research that focuses on a single constituent or uses pairwise comparisons. It is imperative to develop new guidelines for designing the Metaverse to evaluate technological, functional, and experiential variables.

## 6. CONCLUSION

This study examined how visual, environmental, and communication design features of virtual and augmented Metaverse environments affect the tertiary students' experience. Environmental design has the strongest impact on UX through interactivity, cohesiveness, naturalness, and complexity, followed by visual elements that enhance these structures. Communication features did not significantly predict tertiary students' overall UX, suggesting that social connectivity alone does not drive platform adoption. Visual and environmental design were strongly correlated, underscoring the importance of integrated yet non-isolated design decisions. Overall, the three design dimensions explained 47.3% of the variance in UX.

The study's main contribution is the empirical demonstration of a design hierarchy in the VR and AR Metaverse experience. It also quantifies the synergistic relationship between visual and environmental elements, supporting integrated design theory rather than component-level optimization. In practice, the findings suggest that VR/AR development teams should prioritize environmental design, navigation, spatial usability, and interaction support before fine-tuning graphical elements or expanding communication features. Communication tools should be aligned with the specific goals of the application rather than assumed as universal drivers of adoption. A hierarchical approach can guide resource allocation, team composition, including spatial and environmental designers, and quality control criteria for immersive systems.

This study is limited by its associational design, so it cannot make strong claims about cause and effect. The sample is mostly engineering technology and computer science students, thereby limiting the external validity of the findings and constraining their applicability to broader populations, distinct subject cohorts, or diverse sociocultural contexts. The cross-sectional design only captures a short-term participant's experiences. Hence, it does not substantially show how their views change as they gain more experience with the technology. In addition, the model explains only 52.7% of the variance, suggesting that important factors such as technology self-efficacy, personality, intrinsic motivation, and community influences were not fully considered.

Systematic controlled experiments can reveal how design choices in VR and AR affect UX. Communication should be studied in detail because it influences social presence, collaboration, and relationships. Diverse user samples, such as children, older adults, people with disabilities, and people from different cultures, help make the findings more generalizable. Designers should track how beginners and experts improve over time to create better support and customization.

Future research needs multidimensional UX metrics that separate practical, enjoyable, and social aspects of immersive experiences. Emerging technologies such as AI personalization, haptics, brain-computer interfaces, and mixed reality pose new design challenges that require predictive, and theory-driven studies. Strong partnerships between researchers and industry can turn lab findings into real products and make VR and AR experiences more effective, enjoyable, and satisfying for many people. Future work should build on this hierarchy to develop more comprehensive, theory-driven frameworks for designing effective and engaging VR/AR Metaverse experiences.

## FUNDING INFORMATION

No funding was received during the conduct of the study.

## AUTHOR CONTRIBUTIONS STATEMENT

This journal uses the Contributor Roles Taxonomy (CRediT) to recognize individual author contributions, reduce authorship disputes, and facilitate collaboration.

| Name of Author | C | M | So | Va | Fo | I | R | D | O | E | Vi | Su | P | Fu |
|---|---|---|---|---|---|---|---|---|---|---|---|---|---|---|
| Julius G. Garcia | ✓ | ✓ | ✓ | ✓ | ✓ | ✓ | ✓ | ✓ | ✓ | ✓ | ✓ | ✓ | ✓ | |
| Alvie Simonette Q. Alip | | | | ✓ | | ✓ | | ✓ | ✓ | | | | | |

C : **C**onceptualization
M : **M**ethodology
So : **So**ftware
Va : **Va**lidation
Fo : **Fo**rmal analysis
I : **I**nvestigation
R : **R**esources
D : **D**ata Curation
O : Writing - **O**riginal Draft
E : Writing - Review & **E**diting
Vi : **Vi**sualization
Su : **Su**pervision
P : **P**roject administration
Fu : **Fu**nding acquisition

## CONFLICT OF INTEREST STATEMENT

There is no conflict of interest in this study.

## INFORMED CONSENT
Informed consent was obtained from all participants prior to their voluntary involvement.

## DATA AVAILABILITY
The data that support the findings of this study are available on request from the corresponding author, [JGG]. The data, which contain information that could compromise the privacy of research participants, are not publicly available due to certain restrictions.

## BIOGRAPHIES OF AUTHORS

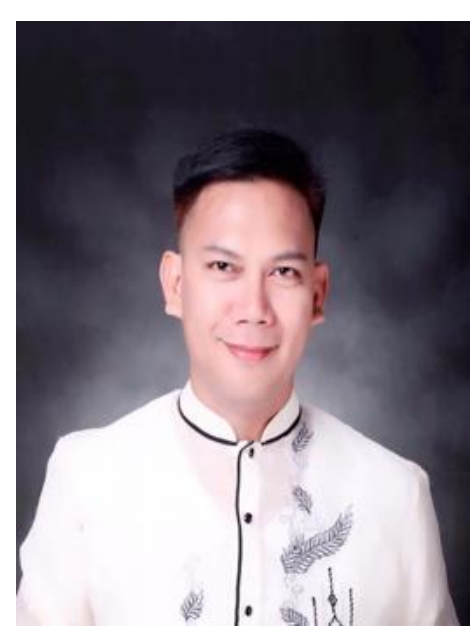

**Julius G. Garcia** is a full professor at the Technological University of the Philippines, Manila, Philippines. He is the head of the International Student Programs and Services at the same university. Currently, he is the publication chair of the World Learning Labs (WILLS) in Kyoto, Japan and one of the board members of the International Exhibition for Young Inventors (IEYI). He has vertically aligned his undergraduate, master's, and doctoral degrees in the field of information technology. He is also a licensed professional teacher. His primary areas of expertise include software development, machine learning, artificial intelligence, project management, and information technology in education. His consulting, involvement in conferences and exhibitions, and publications are all centered on this field. He can be contacted at email: julius.tim.garcia@gmail.com.

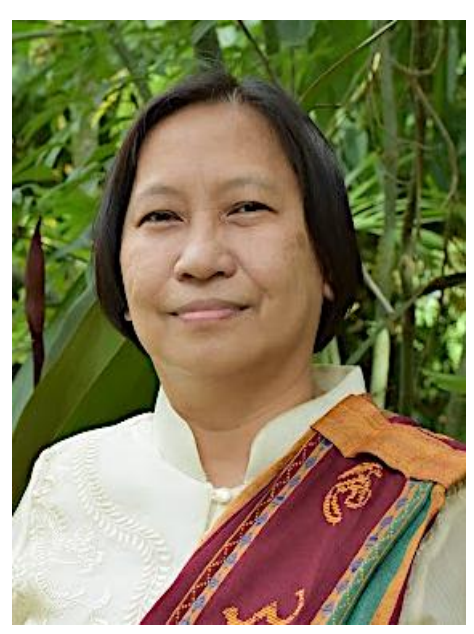

**Alvie Simonette Q. Alip** is a university researcher IV of the office of the vice chancellor for Academic Affairs of the UP Open University. She finished her Bachelor of Science in Biology major in Microbiology from the University of Philippines Los Baños. She took a Master of Arts in Education major in Educational Technology at the University of the Philippines Diliman. She also earned Intensive Course Certificates and Sustainability Science graduate course units from the United Nations University Institute for the Advanced Study of Sustainability. Her main research interests include assessment of student support and digital technologies, educational technologies, education for sustainable development, building sustainable learning communities, resilience in socio-ecological systems, women's role in sustainable development, and university social responsibility. She can be contacted at email: alvie.alip@upou.edu.ph.